\documentclass[aps,prl,twocolumn,preprintnumbers,superscriptaddress,showpacs,nofootinbib,floatfix]{revtex4-1}

\usepackage[normalem]{ulem}

\usepackage[english]{babel}
\usepackage[T1]{fontenc}
\usepackage[colorlinks=true,linkcolor=blue,citecolor=blue,urlcolor=blue]{hyperref}
\usepackage{bm,bbm,amssymb,amsmath}
\usepackage{graphicx}
\usepackage{multirow,dcolumn,booktabs,enumerate}
\usepackage{isotope}
\usepackage{physics}
\usepackage{slashed}
\usepackage{nicefrac}
\usepackage[capitalize]{cleveref}

\allowdisplaybreaks[1]

\usepackage[dvipsnames]{xcolor}
\usepackage{tikz}

\newcommand{\fmiq}{\, \text{fm}^{-3}}

\newcommand{\mev}{\, \text{MeV}}

\newcommand{\EFT}{\ensuremath{\chi\mathrm{EFT}}}

\newcommand{\result}[1]{#1}

\newcommand{\externalresult}[1]{#1}

\newcommand{\prexRskin}{\externalresult{\ensuremath{0.283 \pm 0.071}}}
\newcommand{\prexL}{\externalresult{\ensuremath{106 \pm 37}}}

\newcommand{\CromartieMass}{\externalresult{\ensuremath{2.08 \pm 0.07}}}
\newcommand{\AntoniadisMass}{\externalresult{\ensuremath{2.01 \pm 0.04}}}

\newcommand{\mrgagnPriorEpnm }{\result{\ensuremath{17.5^{+14.6}_{-7.7}}}} 
\newcommand{\mrgagnPriorS    }{\result{\ensuremath{33.3^{+14.7}_{-8.2}}}} 
\newcommand{\mrgagnPriorL    }{\result{\ensuremath{38^{+109}_{-41}}}} 
\newcommand{\mrgagnPriorKsym }{\result{\ensuremath{-255^{+853}_{-566}}}} 
\newcommand{\mrgagnPriorRskin}{\result{\ensuremath{0.14^{+0.19}_{-0.09}}}} 

\newcommand{\mrgagnAstroPostEpnm }{\result{\ensuremath{19.3^{+11.7}_{-8.5}}}} 
\newcommand{\mrgagnAstroPostS    }{\result{\ensuremath{35.1^{+11.6}_{-8.9}}}} 
\newcommand{\mrgagnAstroPostL    }{\result{\ensuremath{58^{+61}_{-56}}}} 
\newcommand{\mrgagnAstroPostKsym }{\result{\ensuremath{-240^{+559}_{-503}}}} 
\newcommand{\mrgagnAstroPostRskin}{\result{\ensuremath{0.19^{+0.12}_{-0.11}}}} 

\newcommand{\mrgagnAstroPREXPostEpnm }{\result{\ensuremath{21.5^{+10.8}_{-8.3}}}} 
\newcommand{\mrgagnAstroPREXPostS    }{\result{\ensuremath{37.3^{+11.8}_{-7.5}}}} 
\newcommand{\mrgagnAstroPREXPostL    }{\result{\ensuremath{80^{+51}_{-46}}}} 
\newcommand{\mrgagnAstroPREXPostKsym }{\result{\ensuremath{-223^{+608}_{-565}}}} 
\newcommand{\mrgagnAstroPREXPostRskin}{\result{\ensuremath{0.23^{+0.10}_{-0.10}}}} 

\newcommand{\mrgeftAstroPostEpnm }{\result{\ensuremath{16.9^{+1.5}_{-1.4}}}} 
\newcommand{\mrgeftAstroPostS    }{\result{\ensuremath{32.7^{+1.9}_{-1.8}}}} 
\newcommand{\mrgeftAstroPostL    }{\result{\ensuremath{49^{+14}_{-15}}}} 
\newcommand{\mrgeftAstroPostKsym }{\result{\ensuremath{-107^{+124}_{-128}}}} 
\newcommand{\mrgeftAstroPostRskin}{\result{\ensuremath{0.17^{+0.04}_{-0.04}}}} 

\newcommand{\mrgeftAstroPostPREXpvalue     }{\result{\ensuremath{12.3\%}}} 
\newcommand{\mrgeftAstroPostPREXhalfpvalue }{\result{\ensuremath{ 0.6\%}}} 

\newcommand{\mrgeftAstroPREXPostEpnm }{\result{\ensuremath{17.1^{+1.5}_{-1.5}}}} 
\newcommand{\mrgeftAstroPREXPostS    }{\result{\ensuremath{33.0^{+2.0}_{-1.8}}}} 
\newcommand{\mrgeftAstroPREXPostL    }{\result{\ensuremath{53^{+14}_{-15}}}} 
\newcommand{\mrgeftAstroPREXPostKsym }{\result{\ensuremath{-91^{+118}_{-130}}}} 
\newcommand{\mrgeftAstroPREXPostRskin}{\result{\ensuremath{0.17^{+0.04}_{-0.04}}}} 

\begin{document}
\preprint{LA-UR-21-20527}


\title{
Astrophysical Constraints on the Symmetry Energy and the Neutron Skin of $^{208}$Pb with Minimal Modeling Assumptions}

\author{Reed Essick}
\email[E-mail:~]{reed.essick@gmail.com}
\affiliation{Perimeter Institute for Theoretical Physics, 31 Caroline Street North, Waterloo, Ontario, Canada, N2L 2Y5}
\affiliation{Kavli Institute for Cosmological Physics, The University of Chicago, Chicago, IL 60637, USA}

\author{Ingo Tews}
\email[E-mail:~]{itews@lanl.gov}
\affiliation{Theoretical Division, Los Alamos National Laboratory, Los Alamos, NM 87545, USA}

\author{Philippe Landry}
\email[E-mail:~]{plandry@fullerton.edu}
\affiliation{Gravitational-Wave Physics \& Astronomy Center, California State University, Fullerton, 800 N State College Blvd, Fullerton, CA 92831}

\author{Achim Schwenk}
\email[E-mail:~]{schwenk@physik.tu-darmstadt.de}
\affiliation{Technische Universit\"at Darmstadt, Department of Physics, 64289 Darmstadt, Germany}
\affiliation{ExtreMe Matter Institute EMMI, GSI Helmholtzzentrum f\"ur Schwerionenforschung GmbH, 64291 Darmstadt, Germany}
\affiliation{Max-Planck-Institut f\"ur Kernphysik, Saupfercheckweg 1, 69117 Heidelberg, Germany}

\begin{abstract}
The symmetry energy and its density dependence are crucial inputs for many nuclear physics and astrophysics applications, as they determine properties ranging from the neutron-skin thickness of nuclei to the crust thickness and the radius of neutron stars.
Recently, PREX-II reported a value of \prexRskin~fm for the neutron-skin thickness of $^{208}$Pb, implying a slope parameter $L=\prexL$~MeV, larger than most ranges obtained from microscopic calculations and other nuclear experiments.
We use a nonparametric equation of state representation based on Gaussian processes to constrain the symmetry energy $S_0$, $L$, and $R_{\rm skin}^{^{208}\text{Pb}}$ directly from observations of neutron stars with minimal modeling assumptions. 
The resulting astrophysical constraints from heavy pulsar masses, LIGO/Virgo, and NICER clearly favor smaller values of the neutron skin and $L$, as well as negative symmetry incompressibilities. Combining astrophysical data with PREX-II and chiral effective field theory constraints yields $S_0 = \mrgeftAstroPREXPostS$~MeV, $L=\mrgeftAstroPREXPostL$~MeV, and $R_{\rm skin}^{^{208}\text{Pb}} = \mrgeftAstroPREXPostRskin$~fm.
\end{abstract}

\maketitle


\textit{Introduction--}
The symmetry energy $S(n)$ is a central quantity in nuclear physics and astrophysics. It characterizes the change in the nuclear-matter energy as the ratio of protons to neutrons is varied and thus impacts, e.g., the neutron-skin thickness of nuclei~\cite{Typel:2001, Vinas:2013hua, Reinhard:2016mdi}, their dipole polarizability~\cite{Tamii:2011pv,Lattimer:2012xj}, and the radius of neutron stars (NSs)~\cite{Lattimer:2000nx, Steiner:2012}.
This information is encoded in the nuclear equation of state (EOS), described by the nucleonic energy per particle, $E_{\rm nuc}/A$, a function of total baryon density $n$ and proton fraction $x=n_p/n$ for proton density $n_p$.
The energy per particle is connected to the bulk properties of atomic nuclei for proton fractions close to $x=1/2$, i.e., symmetric nuclear matter (SNM) with $E_{\rm SNM}/A = (E_{\rm nuc}/A)|_{x=1/2}$.
As the neutron-proton asymmetry increases (or the proton fraction $x$ decreases) the energy per particle increases, reaching a maximum for $x=0$, i.e., pure neutron matter (PNM) with $E_{\rm PNM}/A = (E_{\rm nuc}/A)|_{x=0}$. PNM is closely related to NS matter.
The symmetry energy characterizes the difference between these two systems: 
\begin{equation}
    S(n) = \frac{E_{\rm PNM}}{A}(n)-\frac{E_{\rm SNM}}{A}(n)\,.
    \label{eq:S0}
\end{equation}

Crucial information is encoded in the density dependence of $S(n)$, which is captured by the slope parameter $L$ and the curvature $K_{\rm sym}$ defined at nuclear saturation density, $n_0\approx 0.16 \fmiq$,
\begin{equation}
    L=3 n \left. \frac{\partial S(n)}{\partial n}\right|_{n_0}\,,\quad
    K_{\rm sym}(n)=9 n^2 \left. \frac{\partial^2 S(n)}{\partial n^2}\right|_{n_0}\,.
     \label{eq:L-Ksym}
\end{equation}
As $d (E_\mathrm{SNM}/A)/dn = 0$ at $n_0$, 
$L$ describes the pressure of PNM around $n_0$.
$S_0=S(n_0)$ and $L$ are of great interest to nuclear physics~\cite{Lattimer:2012xj, Tsang:2012se, Huth:2020ozf} and astrophysics~\cite{Oyamatsu:2006vd, Fischer:2013eka, Neill:2020szr}.
Experimental~\cite{Tamii:2011pv, Lattimer:2012xj, RocaMaza:2015, Russotto:2016} and theoretical~\cite{Hebeler:2010jx, Tews:2012fj,  Lonardoni:2019ypg, Drischler:2020hwi} determinations consistently place $S_0$ in the range of $30$--$35$~MeV and $L$ in the range of $30$--$70$~MeV.
Recently, however, the PREX-II experiment reported a new result for the neutron-skin thickness of $^{208}$Pb~\cite{PREXII}, $R_\mathrm{skin}^{^{208}\mathrm{Pb}}$, a quantity strongly correlated with $L$ (see, e.g.,~\cite{Typel:2001, Vinas:2013hua, Reinhard:2016mdi}).
The measurement of $R_\mathrm{skin}^{^{208}\mathrm{Pb}}=\prexRskin$~fm (mean $\pm$ standard deviation), including PREX-I and PREX-II data, led Ref.~\cite{Reed:2021nqk} to conclude that $L=\prexL$~MeV.
This value is larger than previous determinations, and thus presents a challenge to our understanding of nuclear matter, should a high $L$ value be confirmed precisely.

\begin{figure*}[t]
    \begin{tikzpicture}

        \node (figure) {\includegraphics[width=0.60\textwidth, clip=, trim = 0.3cm 0.3cm 0.5cm 0.5cm]{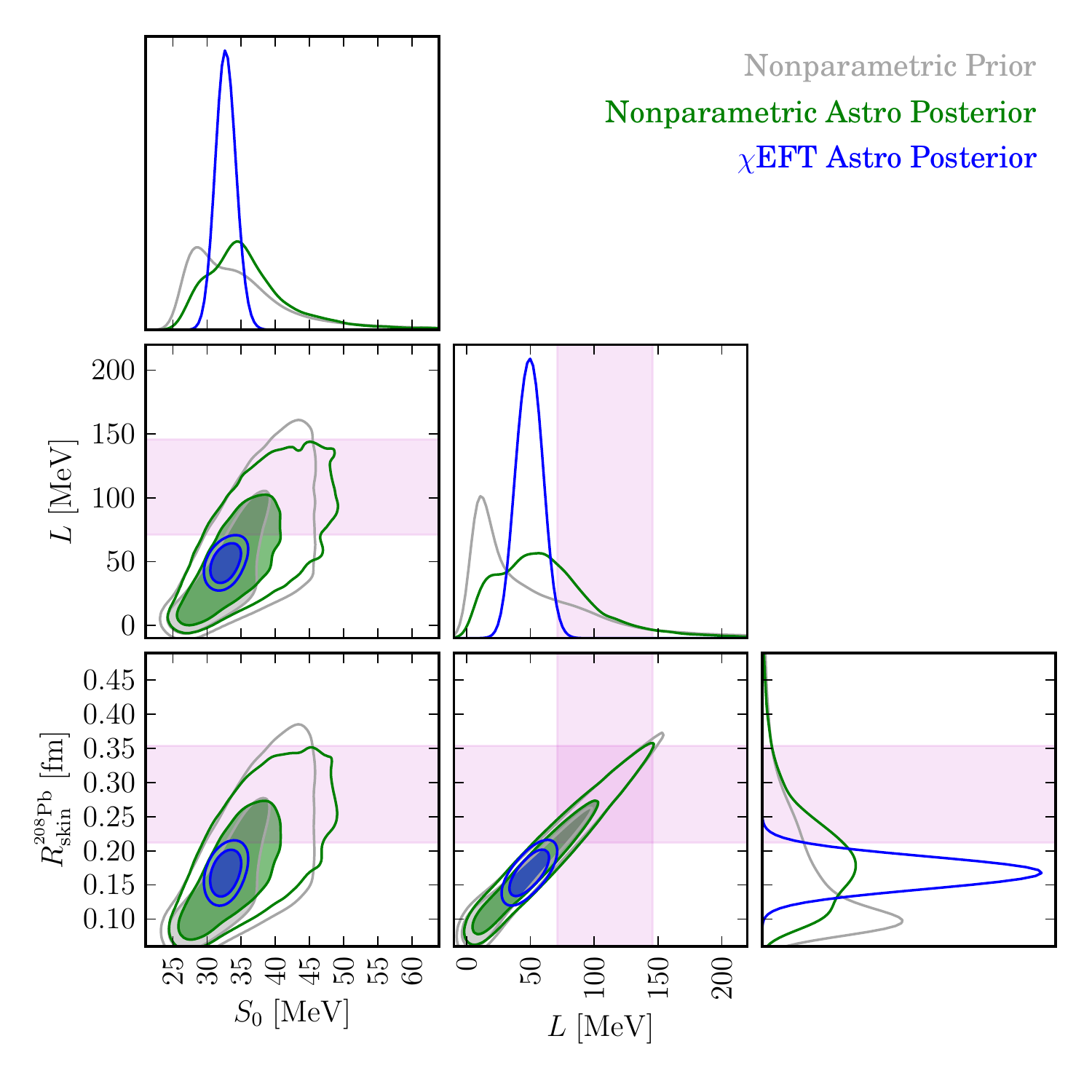}};
        \node (label) [right of=figure, xshift=+3.0cm, yshift=-2.5cm] {\textcolor{red}{PREX-II}};
        
        \node [right of=figure, xshift=+7.0cm, yshift=-3.25cm] {\includegraphics[width=0.18\textwidth, clip=True, trim=1.8cm 0.25cm 0.5cm 0.5cm]{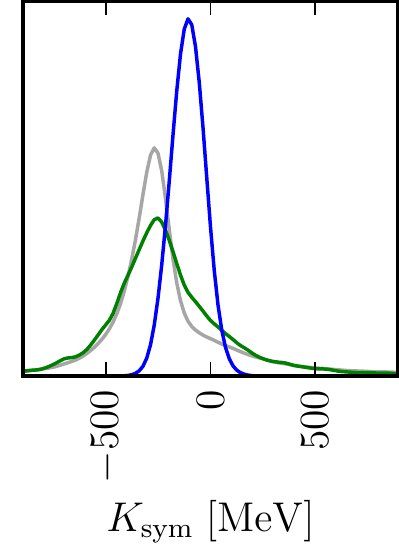}};
    \end{tikzpicture}
    \caption{
        Correlations between the symmetry energy $S_0$, the slope parameter $L$, and the neutron skin thickness of $^{208}$Pb $R_\mathrm{skin}^{^{208}\mathrm{Pb}}$.
        We show the nonparametric prior (\emph{grey}), the nonparametric posterior conditioned on astrophysical observations (\emph{green}), and the nonparametric posterior conditioned on an average over four $\chi$EFT calculations (up to $\approx n_0$) and astrophysical observations (\emph{blue}).
        Joint distributions show the 68\% (\emph{shaded}) and 90\% (\emph{solid lines}) credible regions.
        Shaded bands (\emph{pink}) show the approximate 68\% credible region for parameters constrained by PREX-II: $R_\mathrm{skin}^{^{208}\mathrm{Pb}}$~\cite{PREXII} and the resulting constraints on $L$ using the correlation from Ref.~\cite{Mondal2016}.
        Note how the inclusion of the astrophysical observations shifts the peak in the marginal distributions for $S_0$, $L$, and $R_\mathrm{skin}^{^{208}\mathrm{Pb}}$, a trend that is reinforced by the addition of $\chi$EFT information.
        We also show the one-dimensional marginal distributions for the symmetry incompressibility $K_{\rm sym}$ in a separate panel.
    }
    \label{fig:mrgagn_SLK}
\end{figure*}

In this Letter, we address this question by constraining $S_0$, its density dependence $L$, and $R_\mathrm{skin}^{^{208}\mathrm{Pb}}$ directly from astrophysical observations.
We adopt a nonparametric representation for the EOS~\cite{LandryEssick2019, EssickLandryHolz2020} to minimize the model dependence of the analysis, in contrast to other astrophysical inferences, e.g., Refs.~\cite{AlamAgrawal2016, CarsonSteiner2019, BiswasChar2020, Yue:2021yfx}.
Nonparametric inference allows us to explore a multitude of EOSs that are informed \emph{only} by a NS crust model at densities $n< 0.3 n_0$, where the EOS uncertainty is small, combined with the requirements of causality and thermodynamic stability at higher densities.
Following Ref.~\cite{LandryEssickChatziioannou2020}, the possible EOSs are weighed based on their compatibility with gravitational-wave (GW) and electromagnetic observations of NSs (massive pulsars and X-ray timing with NICER).
By calculating $S_0$, $L$, $K_{\rm sym}$ and $R_\mathrm{skin}^{^{208}\mathrm{Pb}}$ for each of these EOSs, we obtain astrophysically informed posterior distributions for these key nuclear properties.
Furthermore, we study how $L$ and $R_\mathrm{skin}^{^{208}\mathrm{Pb}}$ change as constraints from nuclear theory are included up to progressively higher densities.

\begin{table*}
    \centering
    {\renewcommand{\arraystretch}{1.4}
    \begin{tabular}{@{\extracolsep{0.3cm}} l c c c c c}
        \hline
         &  $E_\mathrm{PNM}/A\ [\mathrm{MeV}]$ &  $S_0\ [\mathrm{MeV}]$ & $L\ [\mathrm{MeV}]$ & $K_\mathrm{sym}\ [\mathrm{MeV}]$ & $R_\mathrm{skin}^{^{208}\mathrm{Pb}}\ [\mathrm{fm}]$ \\
        \hline
        \hline
        Nonparametric Prior
            & \mrgagnPriorEpnm & \mrgagnPriorS & \mrgagnPriorL & \mrgagnPriorKsym & \mrgagnPriorRskin \\ \hline
        Nonparametric Astro Posterior
            & \mrgagnAstroPostEpnm & \mrgagnAstroPostS & \mrgagnAstroPostL & \mrgagnAstroPostKsym & \mrgagnAstroPostRskin \\
        Nonparametric Astro+PREX-II Posterior
            & \mrgagnAstroPREXPostEpnm & \mrgagnAstroPREXPostS & \mrgagnAstroPREXPostL & \mrgagnAstroPREXPostKsym & \mrgagnAstroPREXPostRskin \\
        \hline
        $\chi$EFT Astro Posterior
            & \mrgeftAstroPostEpnm & \mrgeftAstroPostS & \mrgeftAstroPostL & \mrgeftAstroPostKsym & \mrgeftAstroPostRskin \\
        $\chi$EFT Astro+PREX-II Posterior
            & \mrgeftAstroPREXPostEpnm & \mrgeftAstroPREXPostS & \mrgeftAstroPREXPostL & \mrgeftAstroPREXPostKsym & \mrgeftAstroPREXPostRskin \\
        \hline
    \end{tabular}
    }
    \caption{
        Medians and 90\% highest-probability-density credible regions for the studied nuclear properties.
        We compute $R_\mathrm{skin}^{^{208}\mathrm{Pb}}$ from $L$ using the linear fit reported in Ref.~\cite{Mondal2016}, approximating the uncertainty in the fit as described in the text.
    }
    \label{tab:agnostic_credible_regions}
\end{table*}

\textit{Nonparametric inference for the EOS--}
We connect NS observables to $S_0$, $L$, and $K_{\rm sym}$ using a nonparametric representation of the EOS based on Gaussian processes (GPs)~\cite{LandryEssick2019, EssickLandryHolz2020}.
The GPs model the uncertainty in the correlations between the sound speed in $\beta$-equilibrium at different pressures, but do not specify the exact functional form of the EOS, unlike other parameterizations~\cite{ReadLackey2009,Lindblom:2010bb,Lindblom:2012zi,Hebeler:2013nza,Alford:2013aca,RaithelOzel2016,Tews:2018kmu, Tews:2018chv, Greif:2018njt}.
The nonparametric EOSs consequently exhibit a wider range of 
behavior than parametric EOSs, mitigating the impact of modeling assumptions.
The nonparametric EOS inference proceeds through Monte-Carlo sampling from a prior constructed as a mixture of GPs to obtain a large set of EOS realizations.
Each EOS is then compared to astrophysical observations via optimized kernel density estimates (KDEs) of the likelihoods, resulting in a discrete representation of the posterior EOS process as a list of weighted samples (see~\cite{EssickLandryHolz2020, LandryEssickChatziioannou2020} for more details). The posterior probability of a given EOS realization, which we label by its energy density $\varepsilon_\beta$, is calculated as
\begin{equation}
    P(\varepsilon_\beta | \{d\}) \propto P(\varepsilon_\beta) \prod_i P(d_i | \varepsilon_\beta) ,
\end{equation}
where $\{d\} = \{d_1,d_2,\dots\}$ is the set of observations, $P(d_i | \varepsilon_\beta)$ are the corresponding likelihood models, and $P(\varepsilon_\beta)$ is the EOS realization's prior probability. 
The specific likelihoods used in this work are as follows:
(a) Pulsar timing measurements of masses for the two heaviest known NSs (PSR J0740+6620~\cite{CromartieFonseca2020, Fonseca2021}, PSR J0348+0432~\cite{AntoniadisFreire2013}) modeled as Gaussian distributions with means and standard deviations $\CromartieMass\,M_{\odot}$ and $\AntoniadisMass\,M_{\odot}$, respectively; 
(b) GW measurements of masses and tidal deformabilities in the binary NS merger GW170817~\cite{gw170817_props} from Advanced LIGO~\cite{TheLIGOScientific:2014jea} and Virgo~\cite{TheVirgo:2014hva}; and
(c) X-ray pulse-profile measurements of PSR J0030+0451's mass and radius assuming a three-hotspot configuration~\cite{MillerLamb2019} (see also Ref.~\cite{Riley:2019yda}, which yields comparable results~\cite{LandryEssickChatziioannou2020}).

Our basic nonparametric prior can also be conditioned self-consistently on theoretical calculations of the EOS at nuclear densities, while retaining complete model freedom at higher densities~\cite{EssickTewsLandryReddyHolz2020}.
Here we marginalize over the uncertainty bands from four different chiral effective field theory ($\chi$EFT) calculations: quantum Monte Carlo calculations using local $\chi$EFT interactions up to next-to-next-to-leading order (N$^2$LO)~\cite{Lynn:2016}, many-body perturbation theory (MBPT) calculations using nonlocal $\chi$EFT interactions up to next-to-next-to-next-to-leading order (N$^3$LO) of Refs.~\cite{Tews:2012fj,Drischler:2017wtt}, and MBPT calculations with two-nucleon interactions at N$^3$LO and three-nucleon interactions at N$^2$LO (based on a broader range of three-nucleon couplings)~\cite{Hebeler:2009iv,Hebeler:2013nza}.
The resulting marginalized \EFT~band overlaps with results for other realistic Hamiltonians, particularly for Argonne- and Urbana-type interactions~\cite{Gandolfi:2011xu}.
This allows us to account for different nuclear interactions and many-body approaches, increasing the robustness of our results.

To translate the EOS posterior process into distributions for the nuclear physics properties, we establish a probabilistic map from $\varepsilon_\beta$ to $E_\mathrm{PNM}/A$, $S_0$, $L$, and $K_\mathrm{sym}$ (described below). 
Marginalization over the EOS then yields a posterior
\begin{multline}
    P(E_\mathrm{PNM}/A, S_0, L, K_{\rm sym} | \{d\}) = \\ \int \mathcal{D}\varepsilon_\beta\, P(\varepsilon_\beta | \{d\}) P(E_\mathrm{PNM}/A, S_0, L, K_{\rm sym} | \varepsilon_\beta)
\end{multline}
informed by the astrophysical observations. 
Constraints on $R_\mathrm{skin}^{^{208}\mathrm{Pb}}$ are obtained from empirical correlations with $L$~\cite{Mondal2016} calculated from a broad range of nonrelativistic Skyrme and relativistic mean-field density functionals; see also Refs.~\cite{Typel:2001,Reinhard:2016mdi}. To account for the theoretical uncertainty in the fit of Ref.~\cite{Vinas:2013hua} and mitigate its model dependence, we adopt a probabilistic mapping: $P(R_\mathrm{skin}^{^{208}\mathrm{Pb}}|L) = \mathcal{N}(\mu_R, \sigma_R)$ with $\mu_R(L)\,[\mathrm{fm}] = 0.072 + 0.00194 \times (L\,[\mathrm{MeV}])$ and $\sigma_R = 0.0143\,\mathrm{fm}$.

\textit{Reconstructing the symmetry energy--}
Because our nonparametric EOS realizations are not formulated in terms of $S_0$, $L$, or $K_{\rm sym}$, we discuss how to extract the nuclear parameters near $n_0$ directly from the EOS, see the Supplemental Material for more details.
The nonparametric inference provides the individual EOSs in terms of the baryon density $n$ as well as the pressure $p_\beta$ and energy density $\varepsilon_\beta$ in $\beta$-equilibrium.
Each realization is matched to the BPS crust~\cite{BaymPethick1971} around $0.3n_0$.
The choice of a single crust at low densities does not affect our conclusions; see Sec.~V of~\cite{LongManuscript}.
The EOS quantities are related to $E_{\rm nuc}/A$ through
$\varepsilon=n (E_{\rm nuc}/A+m_{\rm N})$
with the average nucleon mass $m_N$.
To reconstruct $E_{\rm nuc}/A$, we correct $\varepsilon_\beta$ by the electron contribution $\varepsilon_{\rm e}$,
\begin{equation}
    \frac{E_{\rm nuc}}{A}(n,x) = \frac{\varepsilon_\beta(n)-\varepsilon_{\rm e}(n,x)}{n} - m_{\rm N}\,. \label{eq:Enuc}
\end{equation}

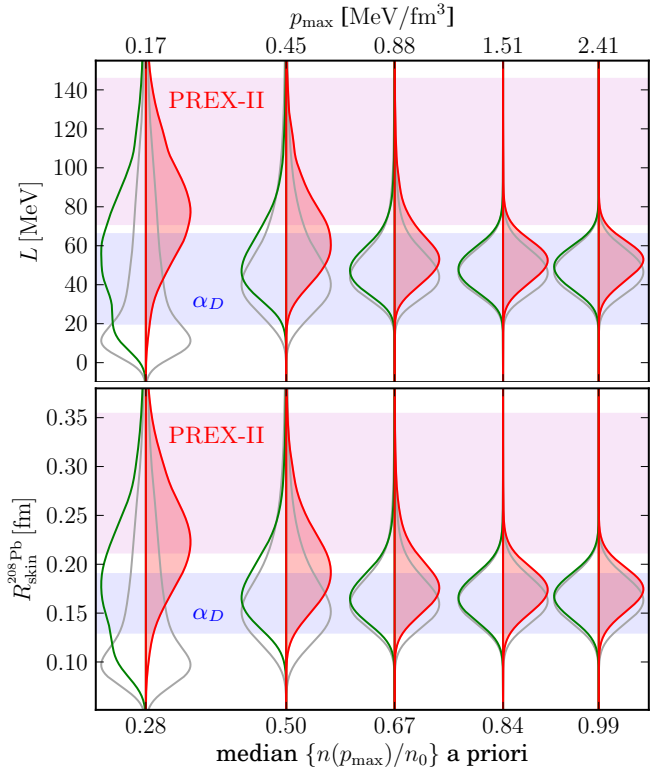
\begin{figure}[t]
    \begin{tikzpicture}
        \node (figure1) {\includegraphics[width=1.0\columnwidth, clip=True, trim=0.6cm 1.5cm 0.6cm 0.0cm]{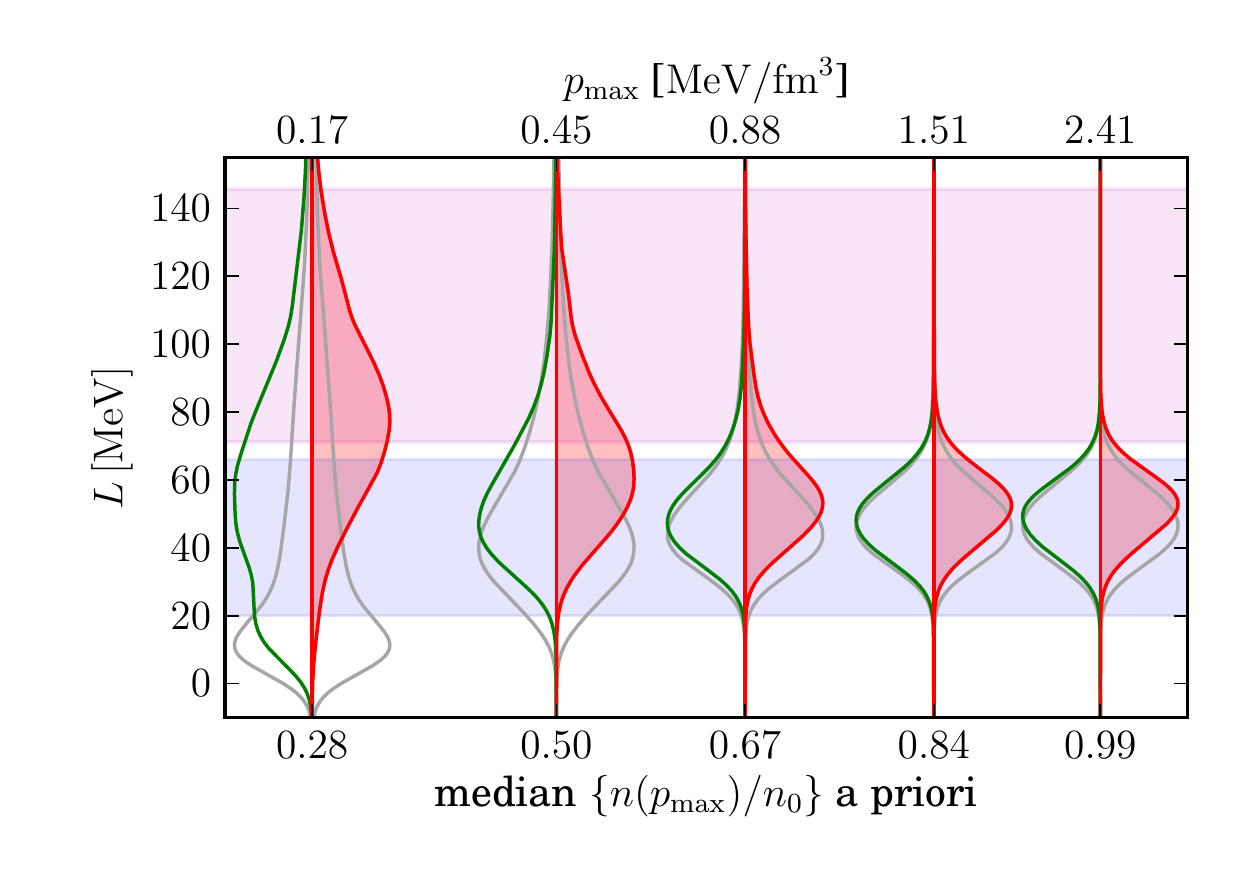}};
        \node (figure2) [below of=figure1, yshift=-4.30cm] {\includegraphics[width=1.0\columnwidth, clip=True, trim=0.6cm 0.5cm 0.6cm 1.5cm]{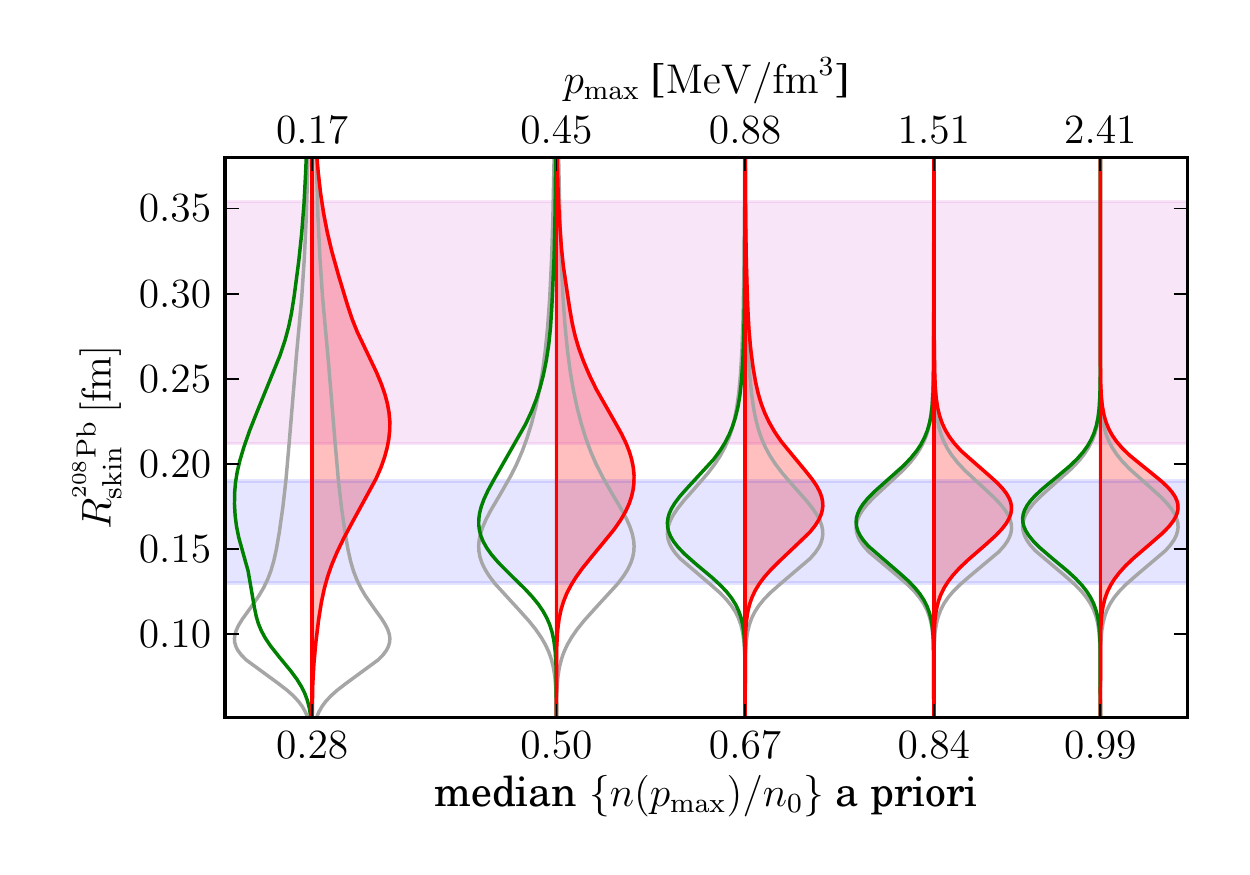}};
        
        \node [right of=figure1, xshift=-2.45cm, yshift=+1.05cm] {\textcolor{red}{PREX-II}};
        \node [right of=figure1, xshift=-2.55cm, yshift=-1.65cm] {\textcolor{blue}{$\alpha_D$}};
        
        \node [right of=figure2, xshift=-2.45cm, yshift=+1.90cm] {\textcolor{red}{PREX-II}};
        \node [right of=figure2, xshift=-2.55cm, yshift=-0.50cm] {\textcolor{blue}{$\alpha_D$}};
    \end{tikzpicture}
    \caption{
        Prior (\emph{gray, unshaded}), Astro posterior (\emph{green}, \emph{left/unshaded}), and Astro+PREX-II posterior (\emph{red, right/shaded}) distributions for $L$ (\emph{top}) and $R_\mathrm{skin}^{^{208}\mathrm{Pb}}$ (\emph{bottom}) as a function of the maximum pressure (\emph{top axis}) or density (\emph{bottom axis}) up to which we trust theoretical nuclear-physics predictions from $\chi$EFT (see text for details).
        Shaded bands show the approximate 68\% credible region from PREX-II~\cite{PREXII} (\emph{pink}) and from Ref.~\cite{RocaMaza:2015} based on the electric dipole polarizability $\alpha_D$ (\emph{light blue}).
    }
    \label{fig:L-R-violin}
\end{figure}

\begin{figure}
    \begin{tikzpicture}
        
        \node (figure1) {\includegraphics[width=1.0\columnwidth, clip=True, trim=0.25cm 0.80cm 0.50cm 0.00cm]{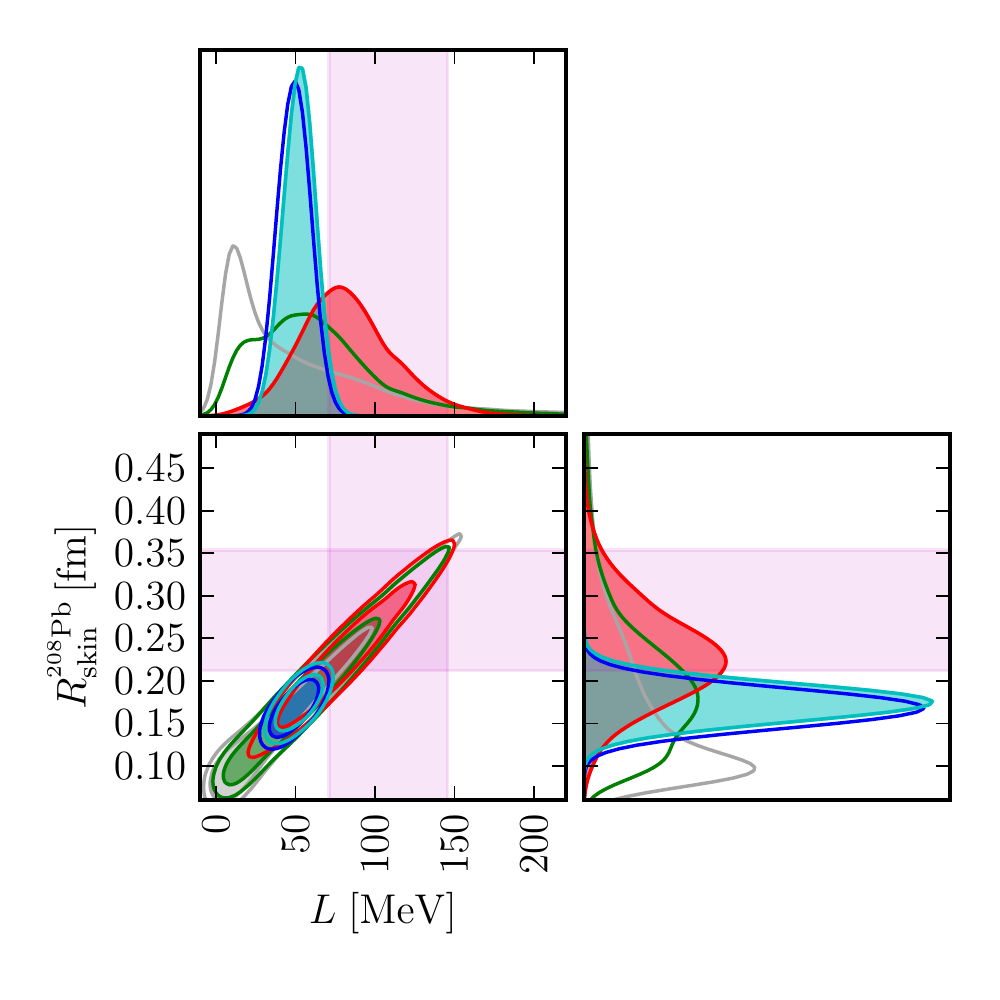}};
        \node (figure2) [below of=figure1, yshift=-4.65cm] {\includegraphics[width=1.0\columnwidth, clip=True, trim=0.25cm 0.50cm 0.50cm 4.30cm]{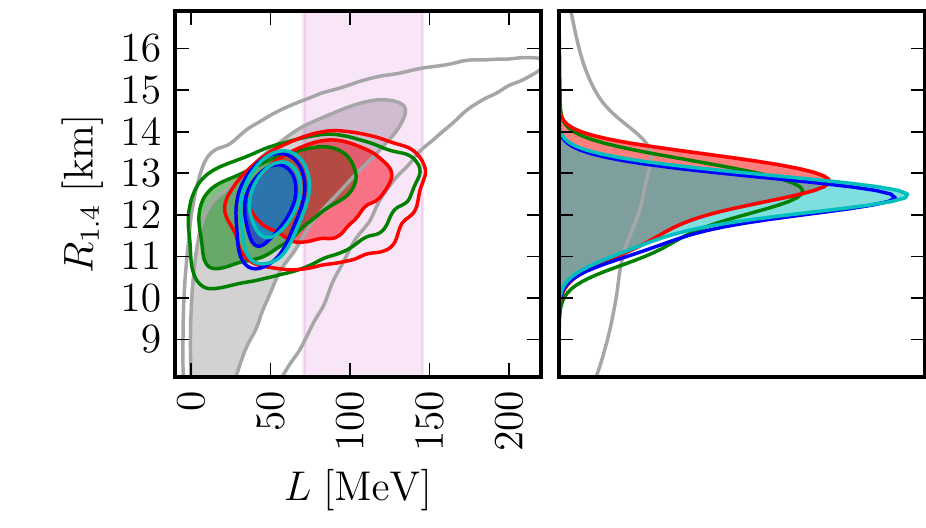}};
        
        \node (label_mrgagn) [right of=figure1, xshift=+2.1cm, yshift=+2.9cm] {\textcolor{red}{Astro+PREX-II}};
        \node [above of=label_mrgagn, xshift=+0.11cm, yshift=-0.63cm] {\textcolor{red}{Nonparametric}};
        \node [below of=label_mrgagn, xshift=+0.53cm, yshift=+0.65cm] {\textcolor{red}{Posterior}};
        
        \node (label_mrgeft) [below of=label_mrgagn, xshift=-0.45cm, yshift=+0.1cm] {\textcolor{cyan}{$\chi$EFT Astro+PREX-II}};
        \node [below of=label_mrgeft, xshift=+1.0cm, yshift=+0.65cm] {\textcolor{cyan}{Posterior}};
        
        \node [right of=figure1, xshift=+2.0cm, yshift=-1.4cm] {\textcolor{red}{PREX-II}};
    \end{tikzpicture}
    \caption{
        Correlations between $R_\mathrm{skin}^{^{208}\mathrm{Pb}}$, $L$, and the radius of a $1.4 M_\odot$ NS, $R_{1.4}$.
        In addition to the priors and posteriors shown in Fig.~\ref{fig:mrgagn_SLK}, we show the nonparametric (\emph{red}) and $\chi$EFT (trusted up to $n_0$; \emph{light blue}) posteriors conditioned on both astrophysical observations and PREX-II.
        Astro+PREX-II posteriors are shaded in the one-dimensional distributions to distinguish them from the Astro-only posteriors.
        Joint distributions show the 68\% (\emph{shaded}) and 90\% (\emph{solid lines}) credible regions.
        Shaded bands (\emph{pink}) show the approximate 68\% credible region from PREX-II.
    }
    \label{fig:L_correlations}
\end{figure}
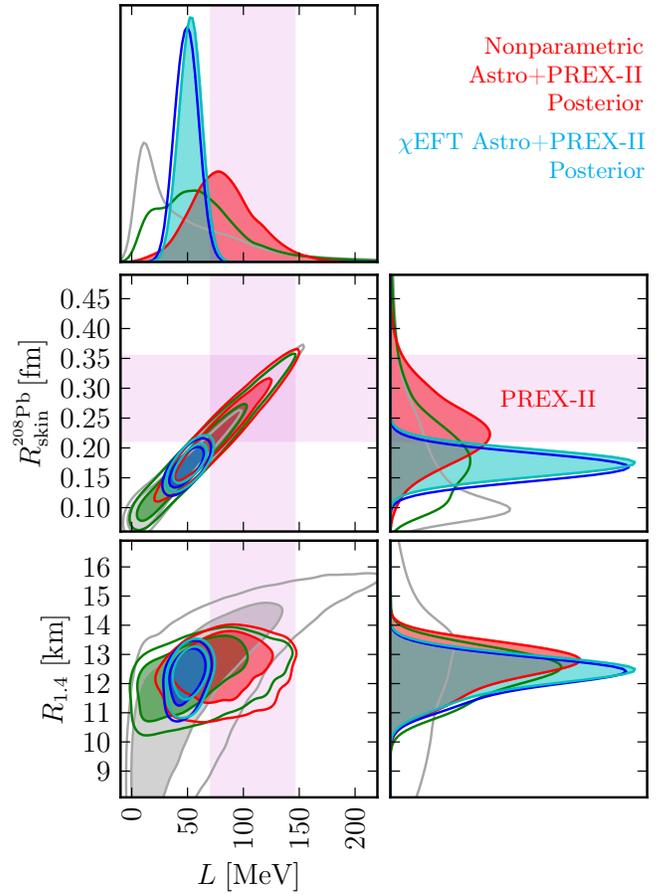

The proton fraction $x(n)$ is unknown and needs to be determined self-consistently for each EOS by enforcing $\beta$-equilibrium, 
$\mu_{\rm n}(n,x)~=~\mu_{\rm p}(n,x)~+~\mu_{\rm e}(n,x)$,
where $\mu_i(n,x)$ is the chemical potential for particle species $i$.
This leads to the condition for $\beta$-equilibrium (see~\cite{Hebeler:2013nza} and the Supplemental Material for details),
\begin{equation}
    0 = m_{\rm n}-m_{\rm p}-\frac{\partial \left(E_{\rm nuc}/A\right)}{\partial x}-\mu_{\rm e}(n,x)\,.
    \label{eq:beta_sol}
\end{equation}
To extract the symmetry energy from each EOS realization, we need to know the dependence of $E_{\rm nuc}/A$ with proton fraction. 
Here, we approximate the $x$ dependence using the standard quadratic expansion,
\begin{align}
    \frac{E_{\rm nuc}}{A}(n,x) & = \frac{E_{\rm SNM}}{A}(n) + S(n) (1-2x)^2  \label{eq:EOS}\,. 
\end{align}
Non-quadratic terms are small at $n_0$ and can be neglected given current EOS uncertainties~\cite{Drischler:2013iza, Somasundaram:2020chb}.
Because we only work around $n_0$, we can characterize the SNM energy using the standard expansion,
\begin{equation}
    \frac{E_{\rm SNM}}{A}(n)=E_0+\frac12 K_0 \left(\frac{n-n_0}{3n_0}\right)^2+\cdots\,,
     \label{eq:SNM}
\end{equation}
where uncertainty in the saturation energy $E_0$, $n_0$, and the incompressibility $K_0$ is based on the empirical ranges from Ref.~\cite{Huth:2020ozf}.
Combining Eqs.~\eqref{eq:S0} and~\eqref{eq:Enuc}--\eqref{eq:SNM}, we find that $\beta$-equilibrium must satisfy
\begin{multline}
    \frac{1-2x_{\beta}}{4}\left(m_p - m_n + \mu_\mathrm{e}(n,x_{\beta})\right) \\
    = \left( \frac{\varepsilon_\beta - \varepsilon_\mathrm{e}(n,x_{\beta})}{n} - m_\mathrm{N} - \frac{E_{\rm SNM}}{A}(n) \right)\,.
    \label{eq:fbeta}
\end{multline}
We use the relations for a relativistic Fermi gas for the electron energy density and chemical potential~\cite{Chamel:2008ca}.

To summarize, given a nonparametric EOS realization and a sample from the empirical distribution for each of the parameters $E_0$, $K_0$, and $n_0$, we reconstruct the proton fraction in $\beta$-equilibrium $x_{\beta}$ self-consistently at each density around nuclear saturation.
We then calculate $E_{\rm PNM}/A$, $S_0$, $L$, and $K_{\rm sym}$ as a function of $n$ and report their values at the reference density $n_0^{(\mathrm{ref})}=0.16\,\mathrm{fm}^{-3}$.
The neutron-skin thickness is estimated via the empirical fit between $R_\mathrm{skin}^{^{208}\mathrm{Pb}}$ and $L$, as discussed above.

\textit{Results and discussion-- }
The constraints on $S_0$, $L$, $K_{\rm sym}$, and $R_\mathrm{skin}^{^{208}{\rm Pb}}$ are shown in Fig.~\ref{fig:mrgagn_SLK}.
We plot the nonparametric prior, the posterior constrained by astrophysical data, and the posterior additionally constrained by the $\chi$EFT calculations up to $n\approx n_0$. 
As our GPs are conditioned on $\chi$EFT up to a maximum pressure ($p_\mathrm{max})$, we report the median density at that pressure (the exact density at $p_\mathrm{max}$ varies due to uncertainty in the EOS from $\chi$EFT).
Prior and posterior credible regions are provided in Tb.~\ref{tab:agnostic_credible_regions}.
We find that the PREX-II result for $R_\mathrm{skin}^{^{208}{\rm Pb}}$ and the extracted range for $L$ of Ref.~\cite{Reed:2021nqk}, \externalresult{$73$}--\externalresult{$147$}~MeV at $1\sigma$, are in mild tension with the GP conditioned on $\chi$EFT calculations up to $n_0$, while the GP conditioned only on astrophysical observations is consistent with both results and cannot resolve any tension due to its large uncertainties. 
However, the Astro-only and $\chi$EFT posteriors peak at similar values for $L$ (\externalresult{$55$}--\externalresult{$65$}~MeV), below the PREX-II result. 
The astrophysical data does not strongly constrain $K_{\rm sym}$, but suggests it is negative.

In Fig.~\ref{fig:L-R-violin}, we show the evolution of our constraints on $L$ and $R_\mathrm{skin}^{^{208}{\rm Pb}}$ as a function of the maximum density up to which we condition on $\chi$EFT, from no conditioning on $\chi$EFT to conditioning on $\chi$EFT up to $n_0$. 
The more we trust $\chi$EFT constraints, the larger the tension with PREX-II results becomes.
We estimate a \mrgeftAstroPostPREXpvalue~probability ($p$-value) that the true $R_\mathrm{skin}^{^{208}{\rm Pb}}$ differs from the PREX-II mean at least as much as the Astro+\EFT~posterior suggests, given the uncertainty in PREX-II's measurement.
However, if a hypothetical experiment confirmed the PREX-II mean with half the uncertainty, this $p$-value would be reduced to \mrgeftAstroPostPREXhalfpvalue.
We also show the estimate for $R_\mathrm{skin}^{^{208}{\rm Pb}}$ obtained from an analysis of dipole polarizability data ($\alpha_D^{^{208}\mathrm{Pb}}$,~\cite{RocaMaza:2015}), which finds $R_\mathrm{skin}^{^{208}{\rm Pb}} = 0.13$--$0.19\,\mathrm{fm}$. 
The latter agrees very well with both the $\chi$EFT results and the nonparametric GP.
See~\cite{LongManuscript} for more comparisons, including joint constraints with both $R_\mathrm{skin}^{^{208}\mathrm{Pb}}$ and $\alpha_D^{^{208}\mathrm{Pb}}$.

In Fig.~\ref{fig:L_correlations}, we present the modeled correlation between $L$ and $R_\mathrm{skin}^{^{208}{\rm Pb}}$ as well as the radius of a $1.4 M_{\odot}$ NS, $R_{1.4}$. 
Besides those shared with Fig.~\ref{fig:mrgagn_SLK}, we show posteriors that are also conditioned on the PREX-II result. 
Even though the results for $L$ and $R_\mathrm{skin}^{^{208}{\rm Pb}}$ are very different for the various constraints, $R_{1.4}$ does not significantly change.
Indeed, the mapping from $L$ to $R_{1.4}$ is broader than often assumed~\cite{Lattimer:2000nx}, and we find that $R_{1.4}$ is nearly independent of our range for $L$.
Hence, the findings of Ref.~\cite{Reed:2021nqk}, indicating that PREX-II requires large radii, include some model dependence. 

\begin{figure}
    \begin{tikzpicture}
        \node (figure) {\includegraphics[width=1.0\columnwidth, clip=True, trim=0.75cm 0.00cm 0.50cm 0.25cm]{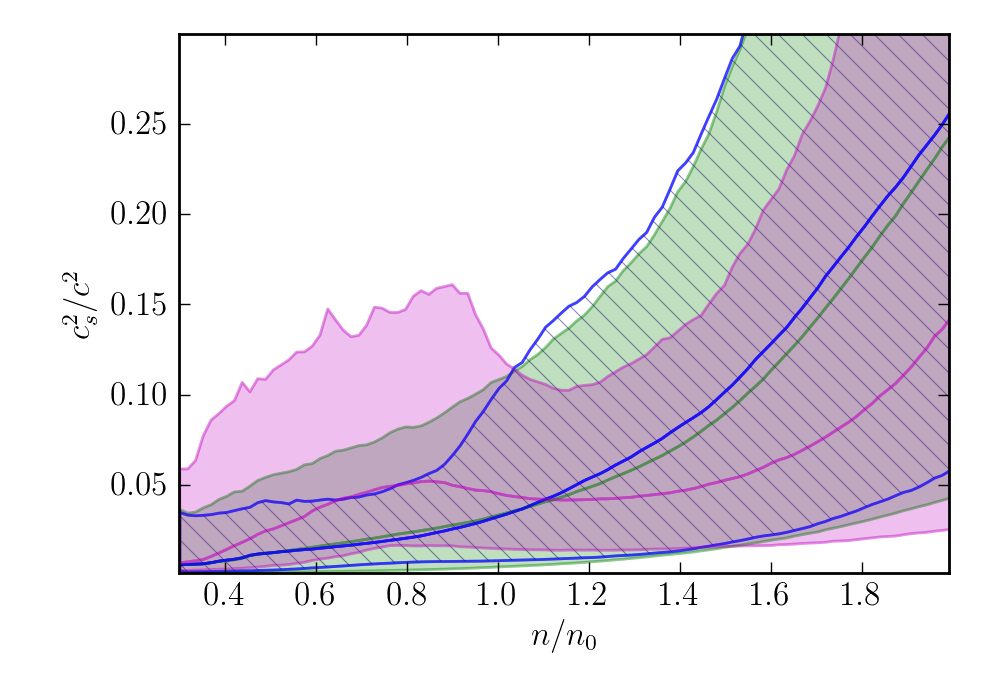}};
        
        \node [left of=figure, xshift=+0.25cm, yshift=+2.5cm] {\textcolor{OliveGreen}{all $L$}};
        \node [left of=figure, xshift=+0.25cm, yshift=+2.0cm] {\textcolor{blue}{$30\,\mathrm{MeV} < L \leq 70\,\mathrm{MeV}$}};
        \node [left of=figure, xshift=+0.25cm, yshift=+1.5cm] {\textcolor{purple}{$100\,\mathrm{MeV} < L$}};
    \end{tikzpicture}
    \caption{
        Median and 90\% one-dimensional symmetric posterior credible regions for $c_s^2$ at each density $n$ with astrophysical observations for all $L$ (\emph{shaded green}), $30\,\mathrm{MeV} < L \leq 70\,\mathrm{MeV}$ (\emph{unshaded blue hatches}), and $100\,\mathrm{MeV} < L$ (\emph{shaded purple}).
    }
    \label{fig:pressure-density with L cuts}
\end{figure}
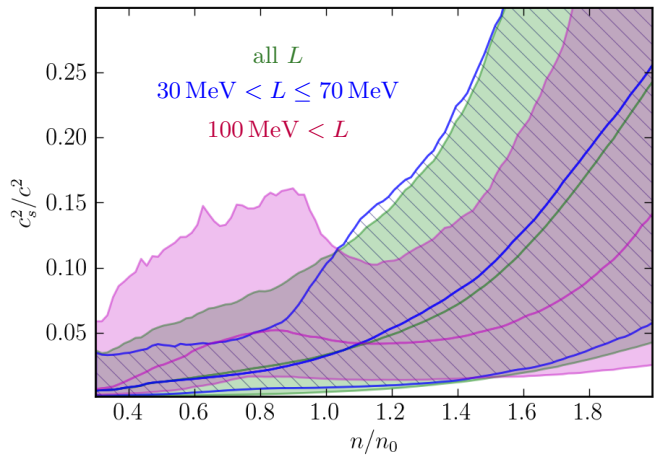

Given the mild tension between the PREX-II value of $R_\mathrm{skin}^{^{208}{\rm Pb}}$ and that inferred from the astrophysical inference with $\chi$EFT information, we investigate what kind of EOS behavior is required to satisfy both the PREX-II and astrophysical constraints.
In Fig.~\ref{fig:pressure-density with L cuts} we show the speed of sound $c_s$ as a function of density for the nonparametric GP conditioned only on astrophysical data for all values of $L$, for $30 \mev <L \leq 70\mev$, and for $L> 100\mev$.
We find that the speed of sound generally increases with density.
However, if we assume $L>100$ MeV, we find a local maximum in the median $c_s(n)$ just below $n_0$, although the uncertainties in $c_s$ are large. 
The reason for this feature is that EOSs that are stiff at low densities (large $L$) need to soften beyond $n_0$ to remain consistent with astrophysical data from GW observations, in particular GW170817. 
Should the PREX-II constraints be confirmed with smaller uncertainty in the future, this might favor the existence of a phase transition between $1$--$2 n_0$. 

In summary, we have used nonparametric GP EOS inference to constrain the symmetry energy, its density dependence, and $R_\mathrm{skin}^{^{208}\mathrm{Pb}}$ directly from astrophysical data, leading to $S_0 = \mrgagnAstroPostS\,\mev$, $L=\mrgagnAstroPostL\,\mev$, and $R_\mathrm{skin}^{^{208}\mathrm{Pb}}=\mrgagnAstroPostRskin\,\mathrm{fm}$. 
Folding in $\chi$EFT constraints reduces these ranges to $S_0 = \mrgeftAstroPostS\,\mev$, $L=\mrgeftAstroPostL\,\mev$, and $R_\mathrm{skin}^{^{208}\mathrm{Pb}}=\mrgeftAstroPostRskin\,\mathrm{fm}$. 
While these results prefer values below the recent PREX-II values~\cite{PREXII, Reed:2021nqk}, in good agreement with other nuclear physics information, the PREX-II uncertainties are still broad and any tension is mild.
Our nonparametric analysis suggests that a $R_\mathrm{skin}^{^{208}\mathrm{Pb}}$ uncertainty of \result{$\pm 0.04\,\mathrm{fm}$} could challenge astrophysical and $\chi$EFT constraints.
Note that the formation of light clusters at the surface of heavy nuclei could affect the extracted L value~\cite{Tanaka:2021oll}.
Finally, our results demonstrate that the correlation between $R_{1.4}$ and $L$ (or $R_\mathrm{skin}^{^{208}\mathrm{Pb}}$) is looser than analyses based on a specific class of EOS models would suggest. 
Extrapolating neutron-skin thickness measurements to NS scales thus requires a careful treatment of systematic EOS model uncertainties.
In particular, the PREX-II result does not require large NS radii.
However, if the high $L$ values of PREX-II persist, this may suggest a peak in the sound speed around saturation density.

\acknowledgments
\textit{Acknowledgements-- }
R.E. was supported by the Perimeter Institute for Theoretical Physics and the Kavli Institute for Cosmological Physics.
Research at Perimeter Institute is supported in part by the Government of Canada through the Department of Innovation, Science and Economic Development Canada and by the Province of Ontario through the Ministry of Colleges and Universities.
The Kavli Institute for Cosmological Physics at the University of Chicago is supported by an endowment from the Kavli Foundation and its founder Fred Kavli.
The work of I.T. was supported by the U.S. Department of Energy, Office of Science, Office of Nuclear Physics, under contract No.~DE-AC52-06NA25396, by the Laboratory Directed Research and Development program of Los Alamos National Laboratory under project numbers 20190617PRD1 and 20190021DR, and by the U.S. Department of Energy, Office of Science, Office of Advanced Scientific Computing Research, Scientific Discovery through Advanced Computing (SciDAC) NUCLEI program.
P.L. is supported by National Science Foundation award PHY-1836734 and by a gift from the Dan Black Family Foundation to the Gravitational-Wave Physics \& Astronomy Center.
The work of A.S.~was supported in part by the Deutsche Forschungsgemeinschaft (DFG, German Research Foundation) -- {Project-ID} 279384907 -- SFB 1245.
This work benefited from discussions within IReNA, which is supported in part by the National Science Foundation under Grant No. OISE-1927130.
The authors also gratefully acknowledge the computational resources provided by the LIGO Laboratory and supported by NSF grants PHY-0757058 and PHY-0823459.
Computational resources have also been provided by the Los Alamos National Laboratory Institutional Computing Program, which is supported by the U.S. Department of Energy National Nuclear Security Administration under Contract No.~89233218CNA000001, and by the National Energy Research Scientific Computing Center (NERSC), which is supported by the U.S. Department of Energy, Office of Science, under contract No.~DE-AC02-05CH11231.


\bibliographystyle{apsrev4-1}
\bibliography{biblio}


\newpage

\begin{center}
\textbf{Supplemental Material}
\end{center}


\section{Extracting Nuclear Parameters from Nonparametric Equations of State in $\beta$-equilibrium}

In the following, we provide detailed information on how to reconstruct the symmetry energy and its density dependence from our nonparametric Gaussian-process (GP) realizations of the equation of state (EOS) in $\beta$-equilibrium.
The nuclear EOS can be described by the nucleonic energy per particle, $E_{\rm nuc}/A(n, x)$, from which $S(n)$ is obtained as
\begin{equation}
    S(n)=\frac{E_{\rm nuc}}{A}\left(n, 0\right)-\frac{E_{\rm nuc}}{A}\left(n, \frac12 \right)\,.
\end{equation}
The nonparametric inference provides the EOS in $\beta$-equilibrium, from which the nucleonic energy per particle must be reconstructed. 

Each EOS realization is represented in terms of the baryon density $n$, the energy density $\varepsilon_{\beta}$ and the pressure $p_{\beta}$ in $\beta$-equilibrium. 
The energy density is related to the energy per particle $E_{\rm nuc}/A$ through
\begin{equation}
    \varepsilon = n \left(\frac{E_{\rm nuc}}{A} +m_{\rm N}  \right)\,,
\end{equation}
where $m_N$ is the average nucleon mass.
Because in $\beta$-equilibrium the energy density contains an electron contribution, we correct for this before we can extract the nucleonic energy per particle,
\begin{equation}
    \frac{E_{\rm nuc}}{A}(n,x) = \frac{\varepsilon_\beta(n)-\varepsilon_{\rm e}(n,x)}{n} - m_{\rm N}\,. \label{eq:supp Enuc}
\end{equation}
We use the relations for a relativistic Fermi gas to decribe the electron contribution~\cite{Chamel:2008ca}, 
\begin{multline}
    \varepsilon_{e}(n_{\rm e}) = \frac{m_{\rm e}^4}{8 \pi^2}
    \left(x_r(2 x_r^2+1)\sqrt{x_r^2+1}\right. \\ 
    \left.-\ln(x_r+\sqrt{x_r^2+1})\right)\,.
    \label{eq:supp el_eps}
\end{multline}
In $\beta$-equilibrium, the electron density equals the proton density, $n_{\rm e} = x n$, and $x_r= k_F / m_{\rm e} =(3 \pi^2 n_{\rm e})^{1/3}/m_{e}$ with the electron mass $m_{\rm e}=0.511$\,MeV.
Around nuclear saturation density, the contribution from muons on the equation of state is negligible (see below for more details).

The proton fraction $x(n)$ is unknown and needs to be determined self-consistently for each GP EOS realization. 
It is constrained by enforcing $\beta$-equilibrium, 
\begin{equation}
    \mu_{\rm n}(n,x)=\mu_{\rm p}(n,x)+\mu_{\rm e}(n,x)\,,
    \label{eq:supp beta_equi}
\end{equation}
where $\mu_{\rm i}(n,x)$ is the chemical potential for particle species $i$.
The electron chemical potential is given by 
\begin{equation}
    \mu_{\rm e}(n_{\rm e}) = \sqrt{(3 \pi^2 n_{\rm e})^{2/3} + m_{\rm e}^2}\,,
    \label{eq:supp el_mu}
\end{equation}
while the neutron and proton chemical potentials $\mu_{\rm n}$ and $\mu_{\rm p}$ in asymmetric nuclear matter are given by
\begin{align}
    & \mu_{\rm p}(n,x)
        = \frac{\partial}{\partial n_{\rm p}} \left[n\left(\frac{E_\mathrm{nuc}}{A} + m_p\right)\right] \nonumber \\[1mm]
        & = n \frac{\partial \left(E_{\rm nuc}/A\right)}{\partial n} + \frac{\partial \left(E_{\rm nuc}/A\right)}{\partial x}(1-x) + \frac{E_{\rm nuc}}{A} + m_{\rm p}\,
\end{align}
and similarly
\begin{align}
    \mu_{\rm n}(n,x) & = n \frac{\partial \left(E_{\rm nuc}/A\right)}{\partial n} - \frac{\partial \left(E_{\rm nuc}/A\right)}{\partial x}x+\frac{E_{\rm nuc}}{A}+m_{\rm n}\,,\label{eq:supp nuc_mu}
\end{align}
with the neutron and proton masses $m_{\rm n}$ and $m_{\rm p}$, respectively.
This leads to the condition for $\beta$-equilibrium,
\begin{equation}
0=m_{\rm n}-m_{\rm p}-\frac{\partial \left(E_{\rm nuc}/A\right)}{\partial x}-\mu_{\rm e}(n,x)\,.
    \label{eq:supp beta_sol}
\end{equation}
We need to know the dependence of $E_{\rm nuc}/A$ on the proton fraction $x$ to extract the symmetry energy. 
In our approach, we approximate the $x$ dependence of the nucleonic energy per particle by using the standard quadratic expansion, 
\begin{align}
    \frac{E_{\rm nuc}}{A}(n,x) & = \frac{E_{\rm SNM}}{A} + S_0 (1-2x)^2  \label{eq:supp EOS}\\
                              & = \frac{E_{\rm SNM}}{A} + \left( \frac{E_{\rm PNM}}{A}-\frac{E_{\rm SNM}}{A} \right) (1-2x)^2\,. \nonumber
\end{align}
Higher-order terms beyond $\mathcal{O}(x^2)$ are expected to be small around $n_0$, and can be safely neglected given current EOS uncertainties~\cite{Drischler:2013iza, Somasundaram:2020chb}.
Then, the derivative of the energy per particle with respect to $x$ is given by
\begin{equation}
    \frac{\partial \left(E_{\rm nuc}/A\right)}{\partial x} =-4\left( \frac{E_{\rm PNM}}{A} - \frac{E_{\rm SNM}}{A}  \right) (1-2x)\,.
    \label{eq:supp deriv_x}
\end{equation}
We describe the SNM energy using the standard expansion around saturation density $n_0$, 
\begin{equation}
    \frac{E_{\rm SNM}}{A}(n)=E_0+\frac12 K_0 \left(\frac{n-n_0}{3n_0}\right)^2+\cdots\,,
     \label{eq:supp SNM}
\end{equation}
where $n_0$, the saturation energy $E_0$, and the incompressibility $K_0$ are constrained empirically.
Here, we use the ranges for these parameters from Ref.~\cite{Huth:2020ozf}:
\begin{align}
    n_0 &= 0.164\pm 0.007 \fmiq\,, \label{eq:supp n0 uncertainty} \\
    E_0 &= -15.86\pm 0.57 \mev\,, \label{eq:supp E0 uncertainty} \\
    K_0 &= 215\pm 40\mev \,, \label{eq:supp K0 uncertainty}
\end{align}
and model our uncertainty in each as a Gaussian with corresponding mean and standard deviation.
Because we only need to extract information on pure neutron matter and the symmetry energy in the immediate proximity of $n_0$ to calculate the empirical parameters $S_0$, $L$, and $K_{\rm sym}$, higher-order terms in the expansion~\eqref{eq:supp SNM} can be neglected (see below for more details).

Hence, we find that $\beta$-equilibrium must satisfy
\begin{multline}
    \frac{1-2x_{\beta}}{4}\Bigl(m_p - m_n + \mu_\mathrm{e}(n,x_{\beta})\Bigr) \\
    = \frac{\varepsilon_\beta(n) - \varepsilon_\mathrm{e}(n,x_{\beta})}{n} - m_\mathrm{N} - \frac{E_{\rm SNM}}{A}(n) \,.
    \label{eq:supp fbeta}
\end{multline}
We solve for $x_{\beta}$ for each GP EOS realization around $n_0$, drawing the parameters $E_0$, $K_0$, and $n_0$ from independent Gaussian models of their empirical distributions (specified above).
In this way, we marginalize over our uncertainty in $E_0$, $K_0$, and $n_0$ via Monte Carlo sampling in the same way that we marginalize over our uncertainty in the EOS by repeated sampling from our GP prior.
With each EOS realization $\varepsilon_\beta$ and set of SNM parameter samples, we can then calculate the PNM energy per particle $E_{\rm PNM}/A(n)$, the symmetry energy $S_0$, its derivative $L$, and its curvature $K_{\rm sym}$ as a function of baryon density $n$. 
We stress that we compute these quantities as functions of density only around $n_0$.
Finally, we report their values at the reference density $n_0^{\mathrm{ref}}=0.16\,\mathrm{fm}^{-3}$.


\subsection{Impact of muons and higher-order terms in $E_\mathrm{SNM}(n)$ on the extracted $S_0$, $L$, and $K_\mathrm{sym}$}

\begin{figure*}[ht]
    \centering
    \includegraphics[width=0.75\textwidth]{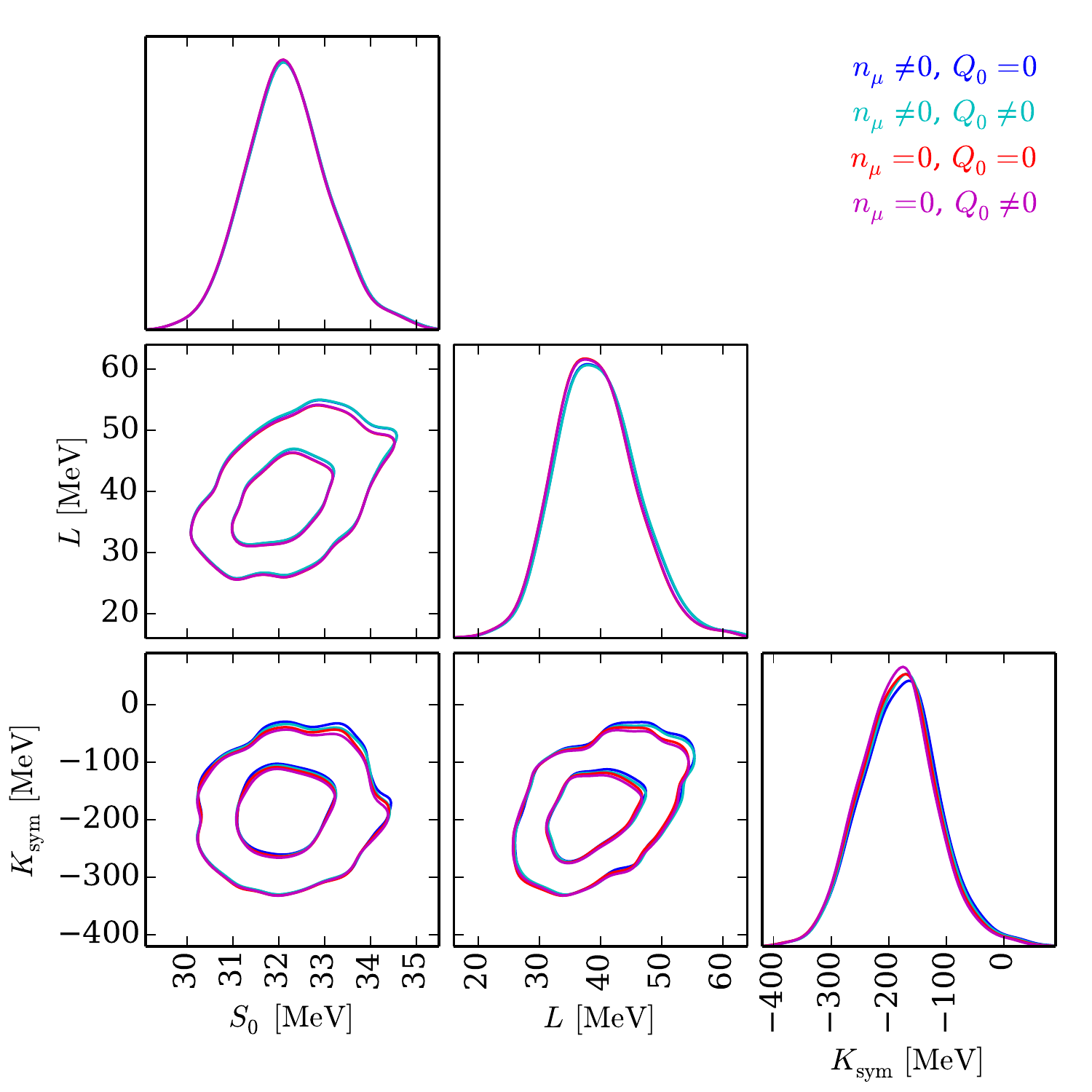}
    \caption{
        Comparison of extracted distributions of $S_0$, $L$, and $K_\mathrm{sym}$ for a set of $\mathcal{O}(10^3)$ EOS realizations based on N$^2$LO QMC calculations~\cite{Lynn:2016}.
        Similar behavior is observed with other EOS priors.
        We report all combinations of including ($n_\mu\neq0$) or neglecting ($n_\mu=0$) muons and truncating the expansion of $E_\mathrm{SNM}$ at second order ($Q_0=0$) or third order ($Q_0\neq0$).
        We see that including muons or higher-order terms in the expansion of $E_\mathrm{SNM}$ has a negligible effect on the extracted symmetry energy and its derivatives.
    }
    \label{fig:supp fancy extraction}
\end{figure*}

Here, we justify two assumptions made within our extraction scheme. Specifically, we consider the impact of muons and of higher-order terms in the expansion of $E_\mathrm{SNM}$ in density around $n_0$.
Figure~\ref{fig:supp fancy extraction} compares the distributions we obtain for $S_0$, $L$, and $K_\mathrm{sym}$ from our prior conditioned on the N$^2$LO QMC calculations~\cite{Lynn:2016}.
Specific values for a few EOS realizations are reported in Table~\ref{tab:supp fancy extraction}.
Similar behavior is observed with the other EOS priors, particularly the other \EFT~calculations considered in this work.

Higher-order terms simply modify our approximation to $E_\mathrm{SNM}(n)$, Eq.~\eqref{eq:supp SNM}.
For example, including a cubic term leads to 
\begin{equation}
    \frac{E_\mathrm{SNM}}{A} = E_0 + \frac{1}{2}K_0\left(\frac{n-n_0}{3n_0}\right)^2 + \frac{1}{6}Q_0\left(\frac{n-n_0}{3n_0}\right)^3 + \cdots
\end{equation}
We approximate the uncertainty in $Q_0$ as Gaussian with a mean of $-300\,\mathrm{MeV}$ and a standard deviation of $100\,\mathrm{MeV}$, based roughly on Table III of~\cite{Somasundaram:2020chb}, drawing a separate realization for each EOS just as we draw realizations for the other parameters describing $E_\mathrm{SNM}$ [Eqs.~(\ref{eq:supp n0 uncertainty}-\ref{eq:supp K0 uncertainty})].

Including muons modifies the condition for $\beta$-equilibrium and the relation between the energy density obtained from our nonparametric EOS realizations and the nuclear energy per particle.
Specifically, charge neutrality implies $n_p = n_e + n_\mu$ and beta equilibrium implies $\mu_e = \mu_\mu$.
These equations can be used to solve for the electron and muon fractions as a function of the proton fraction at each density.
Equation~\eqref{eq:supp Enuc} then becomes
\begin{equation}
    \frac{E_{\rm nuc}}{A}(n,x_e,x_\mu) = \frac{\varepsilon_\beta(n)-\varepsilon_{\rm e}(n,x_e)-\varepsilon_{\mu}(n,x_\mu)}{n} - m_{\rm N}\,, \label{eq:supp Enuc with muons}
\end{equation}
with a corresponding update to Eq.~\eqref{eq:supp fbeta} modeling muons as a degenerate Fermi gas with $m_\mu = 105.658\,\mathrm{MeV}$.
We then solve for the proton fraction in $\beta$-equilibrium as a function of density, from which we extract $S_0$, $L$, and $K_\mathrm{sym}$ as before.

As expected, higher-order terms in the expansion for $E_\mathrm{SNM}$ appear to have almost no effect, because we extract the symmetry energy and its derivatives around $n_0$.
Muons have a larger impact, but the size of the effects is still much smaller than the statistical uncertainty in the prior distribution.
Table~\ref{tab:supp fancy extraction} quantifies the size of the effect for a few example EOS realizations, typically finding the change in $S_0$ is $\mathcal{O}(0.1\%)$, the change in $L$ is $\mathcal{O}(1\%)$, and the change in $K_\mathrm{sym}$ is $\mathcal{O}(5\%)$.

\begin{table*}
    \centering
    {\renewcommand{\arraystretch}{1.4}
    \begin{tabular}{@{\extracolsep{0.6cm}} rrr rrr rrr rrr}
        \hline
        \multicolumn{3}{c}{$n_\mu=0$, $Q_0=0$} & \multicolumn{3}{c}{$n_\mu\neq0$, $Q_0=0$} & \multicolumn{3}{c}{$n_\mu=0$, $Q_0\neq0$} & \multicolumn{3}{c}{$n_\mu\neq0$, $Q_0\neq0$} \\
        \cline{1-3}
        \cline{4-6}
        \cline{7-9}
        \cline{10-12}
        \multicolumn{1}{c}{$S_0$} & \multicolumn{1}{c}{$L$} & \multicolumn{1}{c}{$K_\mathrm{sym}$} & \multicolumn{1}{c}{$S_0$} & \multicolumn{1}{c}{$L$} & \multicolumn{1}{c}{$K_\mathrm{sym}$} & \multicolumn{1}{c}{$S_0$} & \multicolumn{1}{c}{$L$} & \multicolumn{1}{c}{$K_\mathrm{sym}$} & \multicolumn{1}{c}{$S_0$} & \multicolumn{1}{c}{$L$} & \multicolumn{1}{c}{$K_\mathrm{sym}$} \\
        \hline
        \hline
31.56 & 33.4 & $-269$ & 31.57 & 33.7 & $-266$ & 31.56 & 33.4 & $-266$ & 31.57 & 33.7 & $-263$ \\
30.51 & 31.7 & $-103$ & 30.51 & 31.9 & $-99$ & 30.51 & 31.7 & $-103$ & 30.51 & 31.9 & $-99$ \\
33.00 & 46.2 & $-84$ & 33.03 & 47.0 & $-75$ & 33.00 & 46.4 & $-92$ & 33.03 & 47.1 & $-83$ \\
33.32 & 56.4 & $-53$ & 33.35 & 57.4 & $-38$ & 33.32 & 56.5 & $-63$ & 33.35 & 57.5 & $-48$ \\
29.78 & 39.9 & $-200$ & 29.78 & 40.0 & $-196$ & 29.78 & 40.0 & $-209$ & 29.78 & 40.2 & $-204$ \\
        \hline
    \end{tabular}
    \caption{\label{tab:supp fancy extraction}
        Quantification of the impact of different assumptions made while extracting $S_0$, $L$, and $K_\mathrm{sym}$ from a few of the EOS realizations shown in Fig.~\ref{fig:supp fancy extraction}.
        Each row corresponds to a different EOS realization, and the impact of different assumptions can be made by comparing different columns within the same row.
        All values are given in MeV.
    }
    }
\end{table*}


\section{Theoretical Uncertainty in the Map:\\[1mm] $L \rightarrow R_\mathrm{skin}^{^{208}\mathrm{Pb}}$}

\begin{figure*}
    \centering
    \includegraphics[width=0.6\textwidth]{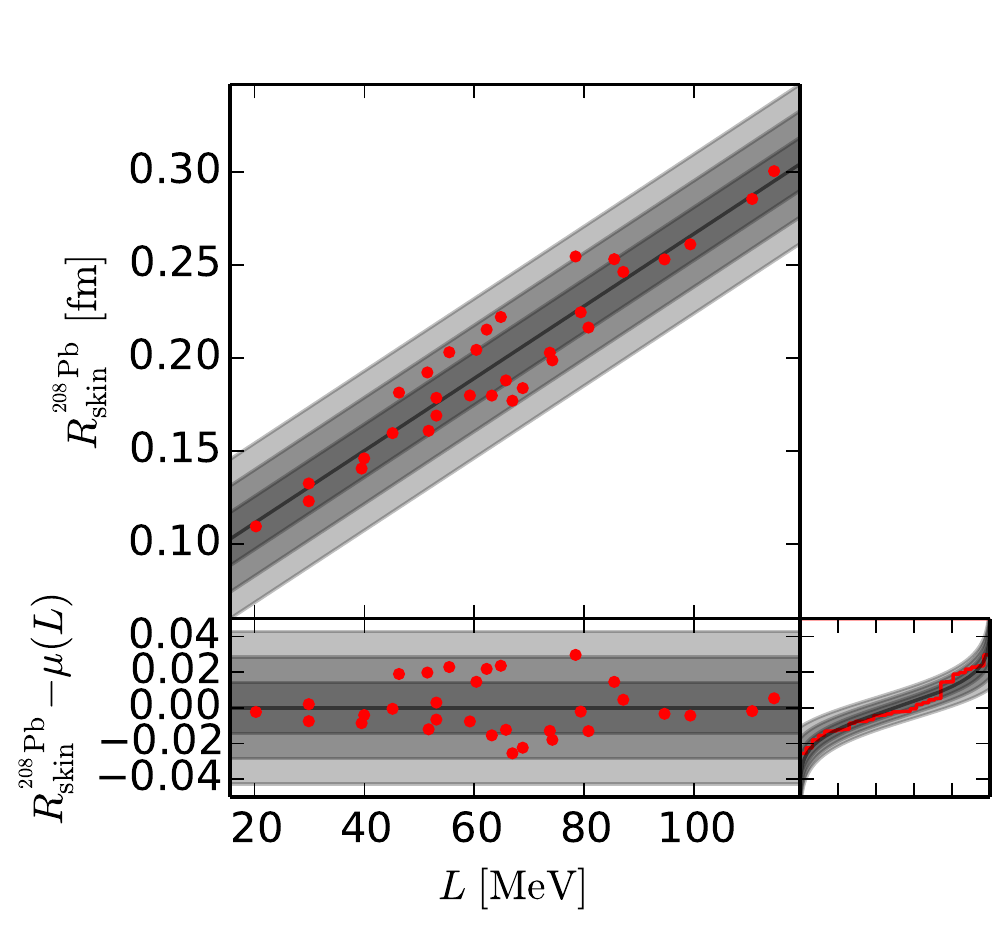}
    \caption{
        \emph{Top panel:} Correlation from Eq.~(\ref{eq:supp uncertainty model}) (\emph{black}) for the 31 models considered in Ref.~\cite{Mondal2016} (\emph{red dots}).
        Shaded regions correspond to 1, 2, and 3-$\sigma$ symmetric credible regions for $R_\mathrm{skin}^{^{208}\mathrm{Pb}}$ at each $L$.
        \emph{Bottom panel:} Residuals about $\mu(L)$ along with a cumulative histogram of the observed distribution of residuals and the Gaussian uncertainty model.
        In the projected histogram, shaded regions demonstrate approximate 1, 2, and 3-$\sigma$ uncertainty bands for the empirical distribution function  (\emph{red line}) given 31 samples from our Gaussian residual distribution (\emph{black}).
        The fact that the empirical distribution closely follows the predicted cumulative distribution demonstrates that our uncertainty model both qualitatively and quantitatively describes the theoretical uncertainty well.
    }
    \label{fig:supp Mondal Uncertainty Model}
\end{figure*}

After extracting $L$ at $n_0^{\mathrm{ref}}$ from our GP EOS representations in $\beta$-equilibrium, we employ an approximate theoretical correlation between $L$ and $R_\mathrm{skin}^{^{208}\mathrm{Pb}}$ in order to compare our results to the PREX experiment.
Although several fits to a variety of theoretical models exist in the literature (see, e.g., Refs.~\cite{Brown:2000pd, Vinas:2013hua, RocaMaza:2015, Mondal2016}), we model the mapping from $L$ to $R_\mathrm{skin}^{^{208}\mathrm{Pb}}$ using the 31 models analyzed in Ref.~\cite{Mondal2016}.
Specifically, we assume Gaussian conditional uncertainty for $R_\mathrm{skin}^{^{208}\mathrm{Pb}}$ given $L$,
\begin{equation}\label{eq:supp uncertainty model}
    p(R_\mathrm{skin}^{^{208}\mathrm{Pb}}|L) = \frac{1}{\sqrt{2\pi \sigma^2}} \exp\left[ - \left(R_\mathrm{skin}^{^{208}\mathrm{Pb}} - \mu(L)\right)^2 \Big/ 2\sigma^2 \right] ,
\end{equation}
where
\begin{align}
    \mu(L)\, [\mathrm{fm}] &= 0.072414 + 0.001943 \times (L\, [\mathrm{MeV}])\, \nonumber \\
    \sigma &= 0.014279\, \mathrm{fm}\,.
\end{align}
Figure~\ref{fig:supp Mondal Uncertainty Model} demonstrates this uncertainty model's behavior.
Although one could approximate the theoretical uncertainty with a more complicated model, including nontrivial changes in $\sigma$ at different $L$, we find that our simple Gaussian model reproduces the quantitative scatter in the residuals very well.
This probabilistic mapping thus explicitly accounts for modeling uncertainty within the correlation between $L$ and $R_\mathrm{skin}^{^{208}\mathrm{Pb}}$.

\end{document}